\documentclass[prl,twocolumn,superscriptaddress,showpacs,amsmath,amssymb]{revtex4-1}
\usepackage{ulem}
\usepackage{mathrsfs}
\usepackage{amsfonts}
\usepackage{amssymb}
\usepackage{amsmath}
\usepackage{bm}
\usepackage{graphicx}
\usepackage{sidecap}
\usepackage{color}
\usepackage[colorlinks=true,citecolor=blue,linkcolor=cyan]{hyperref}
\usepackage{subeqnarray}
\usepackage{verbatim}
\usepackage{ulem}
\usepackage{cases}%the cases package must be put under the amsmath package, still don't known why, just do it
\usepackage{cancel}
\usepackage{framed}
\usepackage{tikz}

\newcommand{\ket}[1]{\vert #1 \rangle}

\begin{document}
%\title{Hybrid Parity-Spin Qubit Operations with Poor-Man's Majorana Modes  }
% \title{Hybrid Operations between  a Spin Qubit and a Poor-Man's Majorana  Parity Qubit}
\title{ Quantum Gates Built on a Spin Qubit and a Kitaev Parity Qubit }
\author{Zhi-Hai Liu}
\email{liuzh@baqis.ac.cn}
\affiliation{Beijing Academy of Quantum Information Sciences, Beijing 100193, China }

\author{Jiang Zhang}
\email{zhangjiang@baqis.ac.cn}
\affiliation{Beijing Academy of Quantum Information Sciences, Beijing 100193, China }

\author{Guilu Long}
\email{gllong@tsinghua.edu.cn}
\affiliation{Beijing Academy of Quantum Information Sciences, Beijing 100193, China }
\affiliation{State Key Laboratory of Low-Dimensional Quantum Physics and
Department of Physics, Tsinghua University, Beijing 100084, China}
\affiliation{Frontier Science Center for Quantum Information, Beijing 100084, China}
\affiliation{Beijing National Research Center for Information Science and Technology, Beijing 100084, China}

\author{H. Q.  Xu}
\email{hqxu@pku.edu.cn}
\affiliation{Beijing Academy of Quantum Information Sciences, Beijing 100193, China }
\affiliation{Beijing Key Laboratory of Quantum Devices, Peking University, Beijing 100871, China}
\begin{abstract}
Spin is typically traced out in the description of quantum-dot-based Kitaev chains to simplify the construction of Majorana fermions. Yet the intrinsic spin structure of poor-man's Majorana modes in minimal Kitaev chains  under finite Zeeman fields offers a natural interface for the Kitaev parity qubit to interact with  other spinful systems.
Here, we establish such a platform to bridge the  parity qubit and a quantum-dot spin qubit, with the effective coupling  governed by the spin-dependent delocalization of the Majorana modes.
Depending on whether the spin qubit is coupled to one or two chains constituting the parity qubit, the parity-spin coupling exhibits distinct forms: an anisotropic parity-conserving   exchange interaction   or a  nontrivial exchange tensor tunable via  the interchain superconducting-phase bias.
Leveraging   fast spin-qubit manipulation, we further demonstrate universal parity-qubit control, high-fidelity qubit-state readout, and entangling operations between the parity and spin qubits.
These results turn the spinful structure of poor-man's Majoranas from a finite-field imperfection into a resource for hybrid quantum control.

\end{abstract}

\date{\today}
\maketitle

\textit{Introduction}.---Semiconductor-superconductor (SM-SC) hybrid quantum dots (QDs) have emerged as a viable platform for realizing Kitaev chains and exploring Majorana fermions~\cite{Bordin2026,Bordin2025,Haaf2025,Bordin2024,Dourado2026,Miles2024,Rodrigo2026,Sau2012,Fulga2012,Liu2025}. Even two-site chains, referred to as minimal Kitaev chains (MKCs), can support poor-man's Majorana bound states (PMMs)~\cite{Zatelli2024,Haaf2024,Tom2023,Leijnse2012}. These states emerge from the competition between elastic cotunneling (ECT) and crossed Andreev reflection (CAR),   mediated by Andreev bound states (ABSs) in the central SM-SC hybrid segment~\cite{Liu2022,Bordin2023,Tsintzis2022,Wang2022,Souto2023,Liu2024}.
Two such MKCs host four PMMs, whose total fermion parity can define a qubit, which we term the Kitaev parity qubit.
To date, theoretical  protocols have  been developed for the qubit initialization, readout,
and   fusion- and braiding-based operations~\cite{Tsintzis2024,Pan2025,Miles2026,Liu2023,Pandey2024,Vimal2026,Liu2026}. Experimentally, single-shot quantum-capacitance measurements have recently enabled real-time parity readout in an MKC~\cite{Loo2026}, followed by the observation of coherent oscillations  of the Kitaev parity qubit~\cite{Zatelli2607}.

Despite these rapid advances, existing control schemes
 largely focus on the    Majorana  occupancy     in MKCs, while the intrinsic spin structure of PMMs is commonly projected out. Indeed, the spinless description of an MKC relies on a sufficiently large Zeeman splitting that energetically separates the relevant spin states~\cite{Luethi2024,Zhang2026}.
In realistic devices, however, the accessible Zeeman field is bounded by the superconducting critical field~\cite{Deng2012,Shen2018,Su2024}.
At finite Zeeman splitting, the accompanying spin-flipping ECT and CAR~\cite{Liu2022,Bordin2023,Wang2022,Liu2024} generally endow PMMs with a nonlocal spin-mixing structure.
Characterizing their intrinsic spin-dependent spatial properties constitutes a necessary step toward determining how Kitaev parity qubits interact with spinful quantum systems, going beyond exploring the parity sector.
In particular,   scalable QD arrays~\cite{Borsoi2024,Zwolak2023,Wang2024,Guido2023} enable  the  parity qubits and (QD-based) spin qubits to be integrated within the same device architecture.
A central question is therefore how the nonlocal spin structure of PMMs determines the interaction between the two qubit species, and whether this interaction can enable universal parity-qubit control, state readout, and entangling operations.

\begin{figure}
  \centering
  \includegraphics[width=0.46\textwidth]{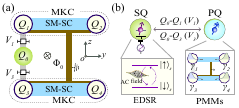}\\
  \caption{(a) SM-SC hybrid QD array integrating a Kitaev parity qubit (PQ) and a spin qubit (SQ) in $Q_{0}$. The PQ is built on the upper and lower MKCs (dashed boxes), with interchain superconducting phase bias controlled by the flux $\Phi_0$. Tunneling between $Q_0$ and $Q_{1,3}$ is tuned by $V_{1,3}$. (b) Left: EDSR control of the SQ encoded by $\ket{\uparrow}_s$ and $\ket{\downarrow}_s$. Right: dominant spatial distributions of the PMMs $\gamma_j$ ($j=1,\ldots,4$) encoding the PQ and the channels mediating PQ-SQ coupling.}
\label{Fig1}
\end{figure}

In this work, we investigate the interaction between a Kitaev parity qubit and a spin qubit in an SM-SC hybrid QD array, as illustrated in Fig.~\ref{Fig1}(a). We first analyze the formation of PMMs in individual MKCs at finite Zeeman splitting and reveal their intrinsic spin-dependent spatial delocalization. By coupling the spin-qubit-hosting QD to one or both constituent chains of the parity qubit, we show that the interdot tunneling   gives rise to an effective exchange interaction   in the strong intradot Coulomb repulsion regime, whose form is determined  by the linkage configuration. Importantly, the exchange strengths are directly connected to the spin-dependent PMM delocalization and, when both chains are coupled, acquire an additional tunability via the interchain superconducting phase bias. Combined with  the well-established fast spin-qubit control~\cite{Noiri2022,Perge2010,Liu2023b,Berg2013,Guido2023}, we further demonstrate universal parity-qubit manipulation, high-fidelity state readout, and entangling gates between the parity and spin qubits. Our work establishes a direct connection between the intrinsic spin structure of PMMs and the controllability of  the Kitaev parity qubit, providing a new route toward hybrid parity-spin architectures.

\textit{Model.}---To construct a Kitaev parity qubit (PQ), two pairs of PMMs are formed in the upper and lower MKCs, consisting of QDs $(Q_1,Q_2)$ and $(Q_3,Q_4)$ interconnected by SM-SC hybrid segments, as enclosed by the dashed boxes in Fig.~\ref{Fig1}(a).  The interchain superconducting phase bias is tuned  by the flux $\Phi_0$ through the grounded loop connecting the two chains. We denote the PMMs in the two chains by $(\gamma_1,\gamma_2)$ and $(\gamma_3,\gamma_4)$  (see Fig.~\ref{Fig1}(b)) and introduce the corresponding nonlocal fermions $f_u= (\gamma_2 + i\gamma_1)/\sqrt{2}$ and $f_w = (\gamma_4 + i\gamma_3)/\sqrt{2}$.
The PQ is defined in a fixed-parity sector of ${\cal N} = f_u^\dagger f_u + f_w^\dagger f_w$.
WLOG, we restrict ourselves to the even-parity subspace $\boldsymbol{\nu}_{e}=\text{Span}\{|00\rangle_{e}, |11\rangle_{e}\}$, where $|mm\rangle_{e}= (f_u^\dagger)^m (f_w^\dagger)^m |0\rangle$ ($m \in \{0,1\}$) and $|0\rangle$ denotes the vacuum state.

The spin qubit (SQ) is encoded in the low-energy Zeeman-split states $\lvert\uparrow\rangle_{s}$ and $\lvert\downarrow\rangle_{s}$ of $Q_0$, located between $Q_1$ and $Q_3$ (see Fig.~\ref{Fig1}(a)).
Unlike the QDs constituting the MKCs, the onsite energy of $Q_0$ is tuned well below the Fermi level into the strong-Coulomb-repulsion regime to stabilize the  qubit (see Fig.~\ref{Fig3}).
Spin-orbit interaction (SOI) further enables rapid SQ manipulation by a resonant AC electric field through electric-dipole spin resonance (EDSR)~\cite{Noiri2022,Perge2010,Liu2023b,Li2013,Golovach2006,Guido2023}, as shown in Fig~\ref{Fig1}(b).
The PQ-SQ coupling is mediated by two independently controlled $Q_{0}$-$Q_{1}$ and $Q_{0}$-$Q_{3}$ tunneling channels, whose barriers are tuned by  $V_{1}$  and $V_{3}$, respectively.

%is encoded by the upper and lower MKCs

\begin{figure}
  \centering
  \includegraphics[width=0.48\textwidth]{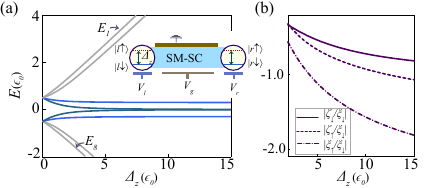}\\
  \caption{(a) BdG spectrum $E_j$ ($E_j\ge E_{j+1}$) versus $\Delta_z$ for  $\alpha_{so}=0.4\pi$, $\epsilon_0=51~\mu\mathrm{eV}$,  and $\Delta_{z} <\Sigma_{0}$. Inset: MKC schematic, with $V_{l/r}$ and $V_g$ the plunger potentials and $\Delta_z$ the Zeeman splitting between the  states  $\lvert l(r)\uparrow\rangle$ and  $\lvert l(r)\downarrow\rangle$. (b) $\log_{10}|\zeta_{\downarrow}/\xi_{\downarrow}|$, $\log_{10}|\zeta_{\uparrow}/\xi_{\downarrow}|$, and $\log_{10}|\xi_{\uparrow}/\xi_{\downarrow}|$ versus $\Delta_z$, obtained from the $E_{4,5}$ eigenstates expanded in the $\boldsymbol{\Upsilon}$ basis.}
\label{Fig2}
\end{figure}

\begin{figure}
   \centering
   \includegraphics[width=0.48\textwidth]{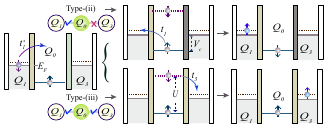}\\
   \caption{ Electron  tunneling around the $Q_1$-$Q_0$-$Q_3$ junction. In Type-(ii), a spin-down electron tunnels out of and back into $Q_1$ via $Q_0$, ending with a spin-up electron in $Q_1$ by  virtual tunneling. In Type-(iii), the electron is transferred from $Q_1$ to $Q_3$ through $Q_0$'s high-energy level, leaving $Q_3$ occupied by a spin-up electron. $E_F$, $V_c$, and $U$ denote the Fermi level and the confinement and Coulomb energies of $Q_0$.}
 \label{Fig3}
 \end{figure}

\textit{Spin-dependent delocalization.}---We first examine the intrinsic spin properties of PMMs formed in an isolated MKC at finite Zeeman splitting. For a magnetic field along the $z$ direction, the QD spin states are separated by the Zeeman splitting $\Delta_{z}$,  as illustrated in the inset of Fig.~\ref{Fig2}(a). The low-energy spin-down states are pinned to the Fermi level by tuning the onsite potentials \(V_{l,r}\).   With   the interdot ECT and
CAR    mediated by ABSs in the central SM-SC hybrid segment~\cite{Liu2022,Bordin2023,Tsintzis2022,Souto2023,Liu2024}, the MKC Hamiltonian is $ H_{ 0}=\sum^{} _{\sigma,\sigma'  }  ( T _{\sigma \sigma^{\prime}_{}} c^{\dagger}_{l\sigma}c^{}_{r\sigma^{\prime}_{}} +  R_{\sigma\sigma^{\prime}_{} }c^{\dagger}_{l\sigma}c^{\dagger}_{r\sigma^{\prime}_{}}+{\rm h.c.}) + \Delta_{z}   (  c^{\dagger}_{l\uparrow}c^{}_{l\uparrow} +  c^{\dagger}_{r\uparrow}c^{}_{r\uparrow} )$, where     $\sigma,\sigma'=\{\uparrow,\downarrow\}$,  and $c_{l\sigma}$ ($c_{r\sigma}$)  annihilates the state $\lvert l\sigma \rangle$ ($\lvert r \sigma\rangle$).   Besides the QD on-site energies,   the spin-dependent ECT and CAR amplitudes  $T_{\sigma\sigma^{\prime}}$  and $R_{\sigma\sigma^{\prime}}$ depend  on  the QD-ABS hopping $t_0$,    the plunger potential $V_g$, and the superconducting gap  (phase)  $\Delta_s$  ($\phi_{s}$)  of the hybrid segment~\cite{Liu2022,Bordin2023,Liu2024,Supplement}.
Incorporating the SOI-induced spin-rotation angle $\alpha_{\rm so}$ across the segment, for example, $T_{\downarrow\downarrow}$ and $R_{\downarrow\downarrow}$ are analytically derived as,
\begin{align}
T_{\downarrow\downarrow} =- \frac{ t^{2}_{0}V_{g}\cos\alpha_{so}}{ V^{2}_{g} +\Delta^{2}_{s}}, ~%\nonumber\\
R_{\downarrow\downarrow} = -i\frac{t^{2}_{0}\Delta_{s} e^{i\phi_{s}}} { V^{2}_{g} +\Delta^{2}_{s}} \sin\alpha_{so}\ .
\end{align}
The ECT-CAR balance $|T_{\downarrow\downarrow}| = |R_{\downarrow\downarrow}|$ imposes   $|V_{g}| =|\tan \alpha_{so}| \Delta_{s}$. PMM formation   further requires tuning  $\Delta_z$.

Figure~\ref{Fig2}(a) shows the BdG spectrum of $H_0$ versus $\Delta_z$, with $E_1 \ge \cdots \ge E_8$. The spectrum is symmetric about zero energy due to particle-hole symmetry. As $\Delta_z$ increases, $E_3$ and $E_6$ asymptotically approach $\pm E_b$, where $E_b = \epsilon_0 |\sin(2\alpha_{\rm so})|/2$ is the magnitude of the balanced $|T_{\downarrow\downarrow}| = |R_{\downarrow\downarrow}|$  with   $\epsilon_0 = t_0^2/\Sigma_0$  and $\Sigma_0 = \sqrt{V_g^2 + \Delta_s^2}$. Concurrently,
$E_{4}$ and $E_{5}$ ($E_{4/5}\equiv \pm E_{c} $) converge to   zero  as $\Delta_z \to \infty$, with $E_{c} \simeq  (\epsilon^{}_{0}/\Delta^{}_{z})^{2} E_{b}[1+\cot^{2}(2\alpha_{so}) ]$~\cite{Supplement}.
Let $f$ and $f^{\dagger}$ denote the quasiparticle  operators associated with the $E_{4}$ and $E_{5}$ modes   at large $\Delta_z$.  A pair of PMMs can then be constructed as $(\gamma, \tilde{\gamma}^{ }  )\equiv  (i f^{\dagger}_{} -i f , f^{\dagger}_{} + f)/\sqrt{2} $. In the Nambu spinor basis   $ \boldsymbol{\Upsilon}=  \{ c_{l\downarrow}, c_{r \downarrow  },  c^{\dagger}_{l\downarrow}, c^{\dagger}_{r \downarrow} , c^{}_{l\uparrow}, c^{}_{r\uparrow} ,c ^{\dagger}_{l\uparrow}, c^{\dagger}_{r \uparrow}    \} ^{\rm T} $,  they take the  form
\begin{align}
\tilde{\gamma} =&\sum_{\sigma=\{\uparrow,\downarrow\}} \eta_{\sigma}\left[ e^{i\frac{\phi_{s}}{2}}( \zeta_{\sigma}e^{i\eta_{\bar{\sigma}}\lambda^{ }_{+}}c^{\dagger}_{l\sigma} +\xi_{\sigma} e^{i\eta_{\bar{\sigma}}\lambda^{ }_{-}}c^{\dagger}_{r\sigma}) +{\rm h.c.} \right],\nonumber \\ \gamma_{} =&\sum_{\sigma } \left[i e^{i\frac{\phi_{s}}{2}}( \xi_{\sigma}e^{i\eta_{\bar{\sigma}}\lambda^{ }_{+}}c^{\dagger}_{l\sigma} +\zeta_{\sigma} e^{i\eta_{\bar{\sigma}}\lambda^{-}_{\sigma}}c^{\dagger}_{r\sigma} )+{\rm h.c.}\right]  ,\label{gama}
\end{align}
where $\xi_\sigma, \zeta_\sigma \in \mathbb{R}$, $\eta_{\downarrow/\uparrow} = \pm 1$, and  $\lambda_{\pm} = (p_{2} \pm p_{1})/2$ , with $p_{1}= \arg( \cos\alpha_{ so}) $ and  $p_{2} = \arg( \sin\alpha_{  so})+\pi/2$ for $V_{g}>0$ . Numerically, \(|\xi_{\downarrow}|\) dominates for  \(\Delta_z> 2 \epsilon_{0}\), indicating that \(\gamma\) and \(\tilde{\gamma}\) are primarily localized in the spin-down states of the left and right QDs, respectively.
The nonvanishing
\begin{align}
 \left(\zeta^{2}_{\downarrow},  \zeta^{2}_{\uparrow}\right) \simeq    \epsilon^{2}_{0}   \xi^{2}_{\downarrow} \Big( \frac{   1 }{ \Delta^{2}_{z}\tan^{2}(2\alpha_{so}) +\epsilon^{2}_{0} }, \frac{ 1}{  \Delta^{2}_{z} + E^{2}_{b} } \Big),
\end{align}
  quantify   the  delocalization of  the PMMs onto the spin-down and spin-up states of their opposite QDs~\cite{Supplement}. Both quantities, together with $\xi_{\uparrow} \propto \epsilon_0/\Delta_z$ describing intradot spin mixing, decrease as $\Delta_z$ increases (see Fig.~\ref{Fig2}(b)). Thus, well-localized and spin-unmixed PMMs are achieved only with sufficiently large $\Delta_z $.

For the two MKCs forming the PQ,   possible differences in $\alpha_{so}$ can be incorporated through the  coefficient sets $(\xi_{\uparrow u},\zeta_{\uparrow u},\xi_{\downarrow u},\zeta_{\downarrow u})$ and $(\xi_{\uparrow w},\zeta_{\uparrow w},\xi_{\downarrow w},\zeta_{\downarrow w})$ for $(\gamma_1,\gamma_2)$ and $(\gamma_3,\gamma_4)$, respectively.  With   the $Q_1$-$Q_0$ and $Q_3$-$Q_0$ channels turned on,  the interchain superconducting phase  bias   also obeys $\phi_{su} - \phi_{sw} = \Phi_0$. Notably, the finite $E_{cu}$ and $E_{cw}$, corresponding to $E_c$ in the two chains,   induce a  nonzero PQ level splitting ($\Omega_e$) at finite $\Delta_{z}$.

\textit{Coupling configurations.}---Depending on whether the $Q_1$-$Q_0$ and $Q_3$-$Q_0$ tunneling channels are open, three distinct PQ-SQ coupling  regimes arise.

\textit{Type-(i)}. For $V_1, V_3 \to \infty$, both channels are closed and the PQ and SQ are decoupled. The  system Hamiltonian is \begin{align}
H^{ d  }_{tot} = \frac{1}{2}(\Omega_e \tau_z  +\Omega_s  \sigma_z)\ ,
\end{align} where $\Omega_e = E_{cu} + E_{cw}$  and $\Omega_s$ is the  SQ   energy splitting,  tunable   via the local field on $Q_0$. $\boldsymbol{\tau}$ and $\boldsymbol{\sigma}$ are the Pauli matrices in the $\boldsymbol{\nu}_e$ subspace and the SQ basis $\{\ket{\uparrow}_s,\ket{\downarrow}_s\}$, respectively.

%For concreteness to values comparable  to  $\Omega_e$

\textit{Type-(ii)}. Lowering $V_{1}$ or $V_{3}$ opens a single tunneling channel.
Concretely, we take the $Q_1$-$Q_0$ channel to be spin conserving and the $Q_3$-$Q_0$ channel to be  spin-flipping to highlight the role of spin.
These processes are captured by $H_{ tun} =   \sum_{\sigma}   ( t_{1} e^{ i \eta_{\sigma}\lambda^{u}_{ +}}  c ^{\dagger}_{1 \sigma}   c _{0 \sigma}+   t_{3} \eta_{\sigma}  e^{ i \eta_{\sigma} \lambda^{ w } _{ + }}  c ^{\dagger}_{3\sigma}    c _{0\bar{\sigma }} +{\rm h.c.})$,  where $c_{k\sigma}$ denotes the spin-state operator in $Q_k$ ($k = 0, \dots, 4$), $t_{1,3}$  vary inversely with $V_{1,3}$~\cite{Liu2021}, and $\lambda^{u,w}_{+}$ generalize $\lambda_+$ for the two MKCs. %are the tunneling amplitudes that
With only the $Q_1$-$Q_0$ channel open ($t_3 = 0$), electron spin tunnels between $Q_0 $ and $Q_1$ (see Fig.~\ref{Fig3}), but the strong intradot Coulomb repulsion $U$~\cite{Shi2012} prevents additional electron injection from $Q_1$ to $Q_0$.  Conversely, escape of the resident SQ carrier from $Q_0$ is hindered by the local confinement potential $V_c$~\cite{Spethmann2022}, with \(V_c \gg \Omega_s\).
Instead of direct tunneling, the PMMs $\gamma_{1,2}$ couple to the SQ  through higher-order virtual tunneling, producing an effective PQ-SQ exchange~\cite{Supplement}. The resulting Hamiltonian is  \begin{align}H ^{ s }_{tot} = \frac{1}{2} (\Omega_e \tau_z  +\Omega_s  \sigma_z)+  J^{u}_{zz}  \sigma_{z} \tau_{z} +  J^{u}_{xz}\sigma_{x}\tau_{z}\ ,\end{align} in which, at moderate $\Delta_z$,
$
 \left(J^{u}_{zz}, J^{u}_{xz}\right)\simeq   \Pi ^{}_{11}\xi_{\downarrow u}  \left(\zeta_{\downarrow u},\zeta_{\uparrow u}\right)
$
with $ \Pi _{\varsigma\varsigma^{\prime}}= -Ut^{}_{\varsigma}t_{\varsigma^{\prime}}/[2V_{c}(U-V_{c})] $ and $\varsigma,\varsigma^{\prime} \in \{1,3\}$. With  this anisotropic exchange, the SQ state flips whereas the PQ does not, since $[H^s_{\rm tot}, \tau_z] = 0$; the same conclusions  hold for other tunneling configurations.  Moreover,  the linear  dependence of ($J^{u}_{zz},J^{u}_{xz}$) on ($\zeta_{\downarrow u},\zeta_{\uparrow u}$ ) elucidates that the PMM delocalization dictates the exchange formation. An analogous result is obtained when only the $Q_0$-$Q_3$ channel is open.

\textit{Type-(iii)}. Lowering both $V_{1}$ and $V_{3}$ opens the two tunneling channels.
In this regime, alongside single-channel high-order virtual tunneling, remote spin transfer between $Q_1$ and $Q_3$ occurs via co-tunneling through $Q_0$ (see Fig.~\ref{Fig3}). In fact, within the BdG framework, electron spin injection into a QD  of  an  MKC    maps  onto hole emission from the same dot and vice versa. Then, the total Hamiltonian for $t_{1,3} \neq 0$ can be obtained as
\begin{align}
H_{\rm tot}^{b}(\Phi_0)
=
\frac{\Omega_s}{2}\sigma_z
+
\frac{\Omega_e}{2}\tau_z
+
\sum_{\alpha,\beta=x,y,z}
J_{\alpha\beta}\,\sigma_\alpha\tau_\beta.\label{Htot}
\end{align}
Here, $J_{zz}= J^{u}_{zz}+J^{w}_{zz}$, $J_{yz}=0$, and $J_{xz}=J^{u}_{xz}+J^{w}_{xz}$, with $(J^{w}_{zz},J^{w}_{xz})\simeq- \Pi_{33} \xi_{\downarrow w} (\zeta_{\downarrow w}, \zeta_{\uparrow w})$.  The remaining exchange terms arise from the interplay between the two channels and can be tuned by $\Phi_0$.  To  first order in $\xi_{\downarrow u}$ and  $\xi_{\downarrow w}$,   the    exchange strengths associated with  PMM delocalization are evaluated as $( J_{xx} , J_{yx}  )\simeq \Pi_{13} h_{\downarrow}  (\cos  \Phi _{v}   , -\sin    \Phi _{v }  )$      and  $ (J_{zx},J_{zy} )\simeq \Pi_{13}   (-h_{ \uparrow}\cos \Phi_{ v} ,  \ell_{\uparrow} \sin \Phi_{ v }   )$,  where $\Phi_{v}= \Phi_{0}/2   $,  $h^{}_{ \sigma}= (\xi_{\downarrow u} \zeta_{\sigma w}-\xi_{\downarrow w}\zeta_{\sigma u} ) $  and  $\ell_{\uparrow} =   (\xi_{\downarrow w}\xi_{\uparrow u} +\xi_{\downarrow u}\xi_{\uparrow w})$~\cite{Supplement}.   By contrast,
$
\left(J_{xy}, J_{yy} \right)\simeq \Pi_{13}  \xi_{\downarrow u} \xi_{\downarrow w}\left(  \sin \Phi_{ v}  ,\cos \Phi_{ v}  \right)
$ are
 insensitive to  the delocalization.
With increasing $\Delta_z$, the drop in $\zeta_{\sigma u}$ and $\zeta_{\sigma w}$ signals the suppression of $J_{xx}$, $J_{yx}$, $J_{zx}$, and $J_{zy}$, so that $J_{xy}$ and $J_{yy}$ dominate at large $\Delta_z$, with $\Phi_0$ controlling their relative strengths.

\begin{figure*}
  \centering
  \includegraphics[width=0.96\textwidth]{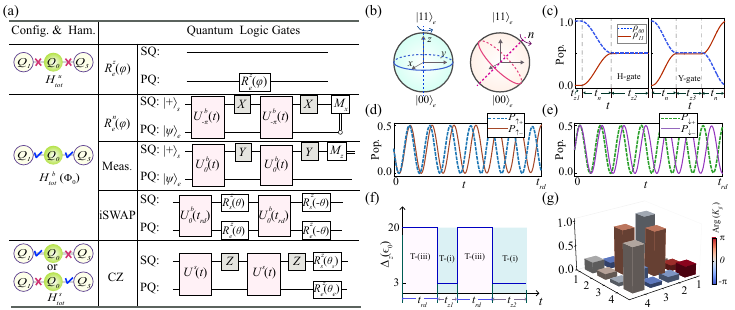}\\
  \caption{(a) Coupling configurations, effective Hamiltonians, and  quantum circuits used for PQ $z$- and $n$-axis rotations, PQ readout, and PQ-SQ iSWAP and CZ gates. (b) PQ rotations about the $\hat z$ and $\hat n$ axes. (c) PQ populations $\rho_{00}$ and $\rho_{11}$ during H and Y gates, with $(t_{z1},t_{z2},t_{z3})\Omega_e=(\pi/2,-\pi/2,-\pi)$   and $t_n\Omega_n=\pi$ $\pmod{4\pi}$. (d,e) Readout dynamics $P_{\sigma\pm}(t)=|\langle\sigma\pm|\Psi(t)\rangle|^2$ for the initial PQ state $(|+\rangle_e+|-\rangle_e)/\sqrt2$, where $|\sigma\pm\rangle=|\sigma\rangle_s\otimes|\pm\rangle_e$. (f) Time sequence of $\Delta_z$ for the iSWAP gate. (g) Magnitudes and phases of the matrix elements $K_{jj'}$ of $\widetilde{U}_{\rm iS}$ after the iSWAP gate, with $j,j' = 1,\dots,4$ indexing the  states in the basis  $\boldsymbol{ \Lambda}$. Calculations use $\alpha_{\rm so}=0.4\pi$ (upper), $0.3\pi$ (lower); $\Delta_{z}=3\epsilon_{0}$ ($\hat z$ rotation),$\Delta_z=8.4\epsilon_0$ ($\hat n$ rotation) and $20\epsilon_0$ (iSWAP); $\epsilon_0=51$ $\mu$eV, $\Omega_{s} =4.0$~$\mu$eV, $t_{1,3}=0.08$ meV,  $V_c=5.0$ meV, $U=9.5$ meV.}
\label{Fig4}
\end{figure*}

\textit{Quantum information processes.}---Quantitatively, analogous to the exchange energy scale of coupled spin qubits~\cite{Matsumoto2026,Reed2016,Noiri2022,Madzik2025}, the coupling strengths in Eq.~(\ref{Htot}) under moderate $\Delta_z$ are typically tens of MHz. This lies below the SQ EDSR frequencies, which can reach hundreds of MHz~\cite{Berg2013} and approach the GHz regime~\cite{Liu2023b}.   The fast EDSR control of the SQ, together with the tunable PQ-SQ exchange, then enables single-qubit operations on the PQ and entangling operations between the two qubits, as summarized in Fig.~\ref{Fig4}(a).

\textit{PQ single-qubit gates}. Arbitrary single-qubit gates can be synthesized from rotations about two nonparallel axes~\cite{Barenco1995}. In the decoupled Type-(i) regime, the finite $\Omega_e$ at moderate $\Delta_z$ drives PQ precession about $z$ axis, $R^z_e(\varphi) = e^{-i\varphi\tau_z/2}$. A second rotation axis is obtained in the  Type-(iii) regime  with $\Phi_{0}=-\pi$ $\pmod{4\pi}$.  Applying EDSR-driven  $x$-axis $\pi$-pulses $R^{x}_{s} (\pi)$  to the SQ in a dynamical decoupling (DD) scheme~\cite{Suter2016,Petta2005,Witzel2007,Bluhm2011,Medford2012,Kuno2026} reduces the Hamiltonian to $\bar{H}^{  b}_{   tot } (-\pi)= [(\Omega_e + 2J_{xz} \sigma_x)\tau_z -2J_{xy}\sigma_x \tau_y]/2$. Preparing the SQ in $\lvert +\rangle_{s} =  (\lvert\uparrow_{ }\rangle_{s} +\lvert\downarrow_{ }\rangle_{s})/\sqrt{2}$ and projecting it back after DD yields an effective PQ rotation about $\hat{n} = (0, -2J_{xy}, \Omega_e + 2J_{xz})/\Omega_{n}$, with $\Omega_{n} =[ 4J^{2}_{xy}+(\Omega_e + 2J_{xz}) ^{2}]^{1/2}$. Because both $\Omega_e$ and $J_{\alpha\beta}$ depend on $\Delta_z$, $\hat{n}$ can be tuned to $(0,1,1)/\sqrt{2}$ (see Fig.~\ref{Fig4}(b)).
The Hadamard (H) and Pauli-$Y$ (Y) gates are then synthesized as
$\text{H}= R^{z}_{e}(-\pi/2) R^{n}_{e}(\pi) R^{z}_{e}(\pi/2)$ and
$\text{Y}=R^{n}_{e}(\pi) R^{z}_{e}(-\pi) R^{n}_{e}(\pi)$,
where $R^{n}_{e}(\varphi)=e^{-i\varphi(\boldsymbol{\tau}\cdot\hat{n})/2}$; other single-qubit gates are discussed in the SM~\cite{Supplement}.
Figure~\ref{Fig4}(c) shows the populations $\rho_{00}$ and $\rho_{11}$ during the H and Y gates for a PQ initialized in $\ket{00}_{e}$.

\textit{PQ state readout}. For $\Phi_0=0$ and large $\Delta_z$, the Type-(iii) coupling is dominated by the $\sigma_y\tau_y$ exchange, allowing the PQ state to be mapped onto the SQ. Write an arbitrary  PQ state as $\lvert\psi\rangle_{e} = b_{+}\lvert +\rangle_{e} +b_{-} \lvert -\rangle_{e}$, with $\tau_y \lvert \pm \rangle_{e} = \pm \lvert \pm \rangle_{e}$, and prepare the SQ in $\lvert +\rangle_{s}$.
To retain only the $J_{yy}$ term, $y$-axis $\pi$-pulses are applied to the SQ to decouple other exchange terms  in $H^{b}_{ tot}(0)$. In addition,  $\Delta_z$ is boosted to ensure $|J_{yy}| \gg \Omega_e $.  Accordingly, over  $t_{rd} = \hbar\pi/(4|J_{yy}|)$, the evolution is well  approximated by $U^{ yy}_{ rd} = e^{i\sigma_y \tau_y \pi/4}$, giving
\begin{equation}
 \lvert\widetilde{\Psi}^{}_{rd}\rangle_{}=  U^{yy}_{rd} \lvert \Psi_{in} \rangle = b_+ \lvert \downarrow \rangle_{s} \lvert + \rangle_{e} + b_- \lvert \uparrow  \rangle _{s} \lvert  - \rangle_{e}\ .
\end{equation}
Then, $|b_{\pm}|^2$ are obtained via projective measurements of the SQ in the basis $\{\lvert\downarrow\rangle_s, \lvert\uparrow\rangle_s\}$. Figures~\ref{Fig4}(d) and \ref{Fig4}(e) show representative readout dynamics $P_{\sigma \pm}(t) = |\langle \sigma \pm  \rvert U^{b}_{0}(t) \lvert \Psi_{in} \rangle_{ }|^2$, where $\lvert \sigma \pm \rangle = \lvert \sigma \rangle_s \otimes \lvert \pm \rangle_e$ and $ U^{b}_{0}(t) $ is the full evolution  operator.  The readout    fidelity $F_{rd} = |\langle \tilde{ \Psi} _{rd} \lvert \Psi_{rd}\rangle|^{2}$ with  $\lvert \Psi_{rd} \rangle =  U^{b}_{0}(t_{rd}) \lvert \Psi_{in} \rangle_{ }  $  exceeds $98\%$.
Combined with the universal PQ rotations above, this projective measurement enables readout in arbitrary bases and full state tomography \cite{Nielsen2000}.

\textit{PQ-SQ  iSWAP  gate}. The iSWAP gate is constructed by combining  the  $\sigma_{y}\tau_{y} $ exchange in Type-(iii)  with $z$ rotations in Type-(i). As in the readout protocol, EDSR-driven $\pi$ pulses and a large $\Delta_z$ isolate the $J_{yy}$ term. Although the iSWAP gate also requires an effective $\sigma_x\tau_x$ interaction, the direct $J_{xx}$ term is small. Instead, the required interaction is generated from $J_{yy}$ by  conjunction  with local $z$ rotations $R^z_{s, e}(\pm\pi/2)$ in the Type-(i) regime. For faster PQ rotation, $\Delta_{z}$ is  reduced to increase $\Omega_{e}$ (see Fig.~\ref{Fig4}(f)).  Then, the resulting iSWAP gate is
\begin{equation}
\widetilde{U}_{\rm iS} =  R^{z}_{\rm t}(-\theta)  \,  U^{b}_{0}(t^{ }_{rd})\, R^{z}_{\rm t}(\theta)  \,  U^{b}_{0}(t^{}_{rd}),
\end{equation}
with $R^{z}_{\rm t}(\pm \theta) =  R^{z}_{s}(\pm\theta) R^{z}_{e} (\pm\theta)$ and $\theta=\pi/2$. Projecting $\widetilde{U}_{\rm iS}$ onto the basis $\boldsymbol{\Lambda} = \{|\uparrow 11\rangle, |\uparrow 00\rangle, |\downarrow 11\rangle, |\downarrow 00\rangle\}$, with $|\sigma m m\rangle \equiv |\sigma\rangle_s \otimes |m m\rangle_e$, Fig.~\ref{Fig4}(g) shows the magnitudes and phases of the matrix elements $K_{j j'}= \langle j \rvert \widetilde{U}_{\rm iS} \lvert  j^{\prime} \rangle$, where $j,j' = 1,\dots,4$ index the basis states in $\boldsymbol{ \Lambda}$ in order.
With respect  to the ideal  iSWAP  gate  $U_{\rm iS}$~\cite{Burkard2023}, the  fidelity of the obtained gate
$F_{\rm iS} =   | \operatorname{tr} (  U _{\rm iS}^\dagger \widetilde{U}_{\rm iS} )  |^{2}/16$ is $90\%$.
The gate fidelity can be further enhanced to $98\%$ by employing the three-segment compensation protocol  as discussed in  SM~\cite{Supplement}, which suppresses the residual noncommuting $\tau_z$ drift.

\textit{PQ-SQ CZ gate}. The Ising-type exchange in Type-(ii), combined with local phase  compensations,   enables a CZ gate~\cite{Bosco2022,Watson2018}.
A moderate reduction of \(\Delta_z\) enhances \(J_{zz}^{u}\), while $z$-axis \(\pi\) pulses on the SQ suppress  the concomitant  $J^{u}_{xz}$ via DD.
Choosing $t_{cz} = \pi\hbar/(4J_{zz}^{u})$ yields $U^{s}(t_{cz}) \simeq e^{-i\pi\sigma_z\tau_z/4}$, up to  local phase rotations. The CZ gate requires further local phase compensation, accomplished by switching to the Type-(i) regime.
The gate is therefore decomposed as
$
\widetilde{U}_{\rm CZ}= R^{z}_{s} ( \theta_{s} ) R^{z}_{e} ( \theta^{ }_{e}  )U^{s}(t_{cz})
$,
with $\theta_{s(e)}=-\theta  -  t_{cz} \Omega_{s(e)}/\hbar $. Including the full time evolution, the fidelity relative to the ideal CZ gate $U_{\rm CZ}$
$F_{\rm cz} =   | \operatorname{tr}( \tilde{U}_{\rm CZ}^\dagger U_{\rm CZ} ) | ^{2}/16$ reaches  $ 99\%$ (see  SM~\cite{Supplement}).

\textit{ Conclusion.}---
In closing, we note prior investigations of hybrid systems consisting of spin and Majorana qubits~\cite{Hoffman2016,Leijnse2011,Li2019}, in which Majorana fermions arise from topological phase transitions  within their respective underlying models. In contrast, the Kitaev parity qubit considered here is made from PMMs realized in MKCs, whose experimental feasibility has been substantially demonstrated~\cite{Zatelli2024,Haaf2024,Tom2023}. Combined with the highly tunable control of these states~\cite{Tom2023,Zatelli2607}, our proposal is therefore within reach of existing experimental capabilities. Beyond idealized theoretical descriptions, our study further reveals the intrinsic spin-dependent delocalization of PMMs under finite-magnetic-field conditions.  We show that this feature can be harnessed as a  resource for  hybrid quantum operations between the parity  and spin qubits.

\begin{acknowledgments}
	\textit{Acknowledgments}.---We thank C. C. Zeng, Y. J. Wu, R. S. Souto, and F. Zatelli for useful discussions.
	Z.-H. L. and H. Q. X. acknowledge support from the National Natural Science Foundation of China
	under Grant Nos. 92565304 and 92165208.
	J. Z. acknowledges support from the National Natural Science Foundation of China
	under Grant No. 12004206.
	G. L. acknowledges support from the National Natural Science Foundation of China
	under Grant Nos. 62471046 and 12634015, the Beijing Advanced Innovation Center for Future Chip (ICFC),
	and the Tsinghua University Initiative Scientific Research Program.
\end{acknowledgments}

\end{document}